\documentclass[12pt,twoside]{article}
\usepackage{amsmath,amsfonts,amssymb}
\usepackage{latexsym}
\usepackage{dcolumn}
\usepackage{graphicx,epsfig}
\usepackage{amsthm}
\usepackage{hyperref}

\begin{document}

\begin{center}
{\Large \bf Singularity Free Rainbow Universe}
\vglue 0.5cm
Barun Majumder\footnote{barunbasanta@iitgn.ac.in}
\vglue 0.6cm
{\small Indian Institute of Technology Gandhinagar \\ Ahmedabad, Gujarat 382424\\ India}
\end{center}
\vspace{.1cm}

\begin{abstract} 
Isotropic quantum cosmological perfect fluid model is studied in the formalism of Rainbow gravity. It is found that the only surviving matter degree of freedom played the role of cosmic time. It is possible to find the wave packet naturally with a suitable choice of the Rainbow functions which resulted from the superposition of the wave functions of the Schr$\ddot{o}$dinger-Wheeler-deWitt equation. The many-worlds interpretation of quantum mechanics is applied to investigate the behavior of the scale factor and the behaviour is found to depend on the operator ordering. It is shown that the model in the Rainbow framework naturally avoids singularity and a bouncing non-singular universe is found.
\vspace{5mm}\newline Keywords: Rainbow universe, quantum cosmology, perfect fluid
\end{abstract}
\vspace{2cm}

\begin{center}
{\bf This essay received an honorable mention in the 2013 Essay Competition of the Gravity Research Foundation}\\ \vspace{2cm}
\end{center}
\newpage

The Planck length $l_P=\sqrt{\frac{\hbar G}{c^3}}$ has some salient features. It has a perfect blend of gravity (G), quantum theory
($\hbar$) and special relativity (c). Whatever new quantum theory of gravity one can formulate with the minimal physical length scale
it should obey special relativity at energy scales much less than the Planck energy. One of the key interest among the candidate theories of quantum gravity is to develop a low energy effective theory consistent with classical general relativity. Recently many issues have been raised from the phenomenological point of view \cite{l01,l011,l11,l2}. Present cosmological observations are expected to observe
leading order effects in $\frac{1}{E_P}$ around the classical background. Theoretically this may mean modification in the dispersion relation which contain energy dependent effects contributed by different orders in $l_P$. Depending on the assumptions related to the Planck scale deformation of Lorentz symmetry the effects can (or cannot) be observed in the observations with ultra high energy cosmic rays and TeV photons \cite{s8,l2,s11}. Even we may observe dependence of the speed of light on energy in gamma-ray bursts \cite{s11}. Quite recently many authors argued that Lorentz symmetry may not be a fundamental symmetry as the construction of classical general relativity does not require Lorentz symmetry. But it is also important that any theory of quantum gravity should respect Lorentz invariance at much lower energies compared to the Planck energy. One of the candidate theory is Doubly Special Relativity which deals with Lorentz invariance very sensitively \cite{l01,l011,l2,l11,s15} in the sense that all inertial observers agree that the speed of light is constant along with Planck energy which is also constant. The transformations act non-linearly on the momentum space introducing terms of the order of $\frac{E}{E_P}$. As an artifact the quadratic invariant is not preserved in momentum space which in turn incorporate corrections to the usual energy momentum dispersion relation. To probe the leading order effect of $l_P$ of the classical spacetime one parameter family of metrics (Rainbow metrics) are proposed with parametrization $\frac{E}{E_P}$. We can consider the notion that spacetime geometry depends on the energy of the particle distorting it \cite{s}. Here $E$ is the scale at which spacetime geometry is probed. In doubly special relativity non-linear Lorentz transformation leads to a modified energy momentum dispersion relation
\begin{equation}
E^2 g_1^2(E/E_P) - p^2 g_2^2 (E/E_P) = m^2
\end{equation}
where $g_1$ and $g_2$ are commonly known as Rainbow functions and $\lim_{E\rightarrow 0} g_{1,2}(E/E_P) =1$. For a doubly general relativity a modified equivalence principle was proposed in \cite{s} which requires that one parameter family of energy dependent orthonormal frame fields describe a one parameter family of energy dependent metrics given by
\begin{equation}
g(E/E_P) = \eta^{ab} e_a(E/E_P) \otimes e_b(E/E_P)
\end{equation}
where $e_0(E/E_P) =\frac{1}{g_1(E/E_P)} \tilde{e}_0$ and $e_i(E/E_P)= \frac{1}{g_2(E/E_P)} \tilde{e}_i$. But in the limit $\frac{E}{E_P}\rightarrow 0$ general relativity must be recovered. With the definition of one parameter
family of energy momentum tensors Einstein's equations are also modified as 
\begin{equation}
G_{\mu \nu}(E/E_P) = 8\pi G(E/E_P)T_{\mu \nu}(E/E_P) + g_{\mu \nu} \Lambda (E/E_P)
\end{equation}
where $G(E/E_P)$ is an effective energy dependent Newton constant expected to satisfy a renormalization group equation. Modified FRW solution is studied in the Rainbow framework in \cite{s} and in \cite{l} the semi-classical Rainbow cosmological model is shown to be singularity free analogous to the bounce in loop quantum cosmology.\par
Inspired by the earlier results in this present work we would like to study the isotropic quantum cosmological perfect fluid model in the Rainbow framework. This is just a toy model to interpret the bounce which is a final outcome of the study in terms of the Rainbow functions. One critical problem in quantum cosmology is certainly that of a suitable choice of time against which the evolution of the universe is investigated. This is because the notion of time has different implications in general relativity and quantum mechanics. If matter is taken as a perfect fluid, the strategy adopted by Schutz \cite{sch} becomes extremely useful as a set of canonical transformations leads to one conjugate momentum associated with the fluid giving a linear contribution to the Hamiltonian. The corresponding fluid variable thus qualifies to play the role of time in the relevant Schr$\ddot{o}$dinger equation. In an ever expanding model, the fluid density has a monotonic temporal behavior, the time orientability is thus ensured. Schutz's formalism has been extensively used by many authors for isotropic \cite{iso} and anisotropic \cite{aniso,nb} cosmological models. Matter content is taken as a perfect fluid with an equation of state $p=\alpha \rho$. For the Rainbow isotropic model we found finite norm wave packet solutions of the Wheeler-deWitt equations. One important finding is that singularity free cosmological model could be constructed even without violating the energy conditions with a suitable choice of $g_2(E/E_P)$. \par
The relevant action for gravity with a perfect fluid can be written as
\begin{equation}
S = \int_M d^4 x \sqrt{-g}~ R + 2 \int_{\partial M} d^3 x \sqrt{h}~ h_{ab}~K^{ab} + \int_M d^4 x\sqrt{-g}~ P
\end{equation}
where $h_{ab}$ is the induced metric over three dimensional spatial hypersurface which is the boundary $\partial M$ of the four dimensional manifold $M$ and $K^{ab}$ is the extrinsic curvature. Here units are so chosen that $16\pi G=\hbar=1$. $P$ is the pressure due to the perfect fluid which satisfies the equation of state $P=\alpha \rho$ with $\alpha <1$. This restriction stems from the consideration that sound waves cannot propagate faster than light. The Rainbow FRW metric for a homogeneous and isotropic universe can be written as 
\begin{equation}
dS^2 = \frac{N^2(t)}{g_1^2(E/E_P)} dt^2 - \frac{a^2(t)}{g_2^2(E/E_P)} \left[ \frac{dr^2}{1-kr^2} + r^2 d\vartheta^2 + r^2 \sin^2 \vartheta
d\varphi^2 \right]
\end{equation}
where $N$ is the lapse function and $k=0,1,-1$ correspond to flat, closed and open universe respectively. $g_1$ and $g_2$ are the Rainbow functions. Let us first calculate the Hamiltonian for the fluid part. In Schutz's formalism \cite{sch} the fluid's four velocity can be expressed in terms of six potentials. However two of them are connected with rotation. FRW type models permit timelike geodesics which are hypersurface orthogonal, the rotation tensor $\omega_{\mu \nu}$ vanishes and one can write the four velocity in terms of only four independent potentials as
\begin{equation}
u_{\nu} = \frac{1}{h} (\epsilon_{,\nu} + \theta S_{,\nu}) ~~.
\end{equation}
Here $n, S, \epsilon$ and $\theta$ are the velocity potentials having their own evolution equations, where the potentials connected with vorticity are dropped. The four velocity is normalized as $u_{\nu}u^{\nu} =1$. Although the physical identification of velocity potentials are irrelevant for the formulation, $h$ and $S$ can be identified with the specific enthalpy and specific entropy
respectively. This identification facilitates the representation of fluid parameters in terms of thermodynamic quantities. Using the thermodynamic relations of \cite{sch} the Lagrangian for the perfect fluid can be written as
\begin{equation}
L_f = \frac{g_1^{\frac{1}{\alpha}}(E/E_P)}{g_2^3(E/E_P)} ~a^3 N^{-\frac{1}{\alpha}}~ \frac{\alpha (\dot{\epsilon}+\theta \dot{S})^{1+\frac{1}{\alpha}}}{(1+\alpha)^{1+\frac{1}{\alpha}}}~ e^{-\frac{S}{\alpha}} ~~.
\end{equation}
As $h>0$ so $(\dot{\epsilon}+\theta \dot{S})>0$. We can write the Hamiltonian of the perfect fluid in final form as
\begin{equation}
H_f = \frac{N g_2^{3\alpha}(E/E_P)}{g_1(E/E_P)} \frac{P_T}{a^{3\alpha}} ~~.
\end{equation}
Here we have used the following canonical transformations of \cite{nb12}
\begin{equation}
T=-P_S e^{-S}P_{\epsilon}^{-\alpha-1} ~~~~~~~~~~~\text{and}~~~~~~~~~~~ P_T = P_{\epsilon}^{\alpha+1} e^S
\end{equation}
to write the Hamiltonian in the simple form (linear in $P_T$) where $P_{\epsilon}=\frac{\partial L_f}{\partial \dot{\epsilon}}$, $P_{S}=\frac{\partial L_f}{\partial \dot{S}}$ and $P_S = \theta P_{\epsilon}$. The advantage of using this method, i.e., using canonical transformations, is that we could find a set of variables where the system of equations is more tractable, while
the Hamiltonian structure of the system remains intact. The super Hamiltonian for the minisuperspace of this model can now be written as
\begin{equation}
H=H_g + H_f = -\frac{N}{g_1(E/E_P)} \left[\frac{g_2^3(E/E_P)}{24} \frac{p_a^2}{a} + \frac{6ka}{g_2(E/E_P)} - g_2^{3\alpha}(E/E_P)~ \frac{P_T}{a^{3\alpha}} \right] ~~
\end{equation}
where we have incorporated the Hamiltonian for the geometry sector. Here $N$ is the Lagrange multiplier taking care of the classical constraint equation $H=0$. In this equation $T=t$ may play the role of cosmic time if we choose the gauge $N=\frac{g_1(E/E_P)}{g_2^{3\alpha}(E/E_P)}a^{3\alpha}$ and this follow from the classical equation as $\dot{T}=\{T,H\} = \frac{Ng_2^{3\alpha}(E/E_P)}{g_1(E/E_P)a^{3\alpha}}$. Using the usual quantization procedure \cite{nb13} we write the Schr$\ddot{o}$dinger-Wheeler-deWitt equation for our super Hamiltonian with the ansatz that the super Hamiltonian operator annihilates the wave function
\begin{equation}
\frac{\partial^2 \Psi(a,T)}{\partial a^2} - \frac{q}{a} \frac{\partial \Psi (a,T)}{\partial a} - \frac{144 k a^2}{g_2^4(E/E_P)}
+ i 24 g_2^{3(\alpha -1)}(E/E_P)a^{1-3\alpha} \frac{\partial \Psi (a,T)}{\partial T} = 0
\end{equation}
with $p_a \rightarrow -i\frac{\partial}{\partial a}$, $P_T \rightarrow i\frac{\partial}{\partial T}$ and $q$ is the operator ordering parameter. Here we can see that Rainbow function $g_1(E/E_P)$ do not take part in the Schr$\ddot{o}$dinger-Wheeler-deWitt equation just like $N$. This can be seen as a scaling of the lapse function $N$ by $\frac{1}{g_1(E/E_P)}$. In order to solve for the wave function $\Psi$ we employ a separation of variables as
\begin{equation}
\Psi (a,T) = e^{-iET} \phi (a)
\end{equation}
to get for flat case $k=0$
\begin{equation}
\label{phieq}
\frac{\partial^2 \phi}{\partial a^2} - \frac{q}{a} \frac{\partial \phi}{\partial a} + 24 E g_2^{3(\alpha -1)}(E/E_P)a^{1-3\alpha} \phi = 0
\end{equation}
Here we would like to interpret $E$ as the scale at which the minisuperspace (in our case) is probed. The motivation for this comes from the fact that the canonically conjugate momenta ($P_T$) of $T$ is linear in the super Hamiltonian. In analogy to quantum mechanics this is equivalent to the time independent Schr$\ddot{o}$dinger equation $\hat{H}\psi = E\psi$. Further we would like to consider the fact that the perfect fluid is composed of baryons which can undergo transmutation but the baryon number is conserved \cite{sch}. Following the proposal of \cite{s} here we would like to emphasize that $E$ is the total energy of the system of baryons
that is measured by a freely falling inertial observer. \par
The solution of (\ref{phieq}) is known and we can construct a regular wave packet superposing these solutions capable of describing physical states. The construction of the wave packet can be expressed as
\begin{equation}
\label{wp1}
\Psi_{wp} = C_3~ a^{\frac{q+1}{2}} \int_0^{\infty} g_2^{-\frac{(q+1)}{2}}(E/E_P)~r^{\nu +1}~ e^{-i\frac{3}{32}(1-\alpha)^2 T r^2}
~J_{\nu}\Big\{r g_2^{-3\frac{(1-\alpha)}{2}} a^{\frac{3(1-\alpha)}{2}}\Big\} dr
\end{equation}
where $r=\frac{4\sqrt{6E}}{3(1-\alpha)}$, $C_3 = \frac{3(1-\alpha)^2}{16}[54(1-\alpha)^2]^{\frac{q+1}{6(1-\alpha)}}$ and $\nu = \frac{q+1}{3(1-\alpha)}$. Now let us choose a form of $g_2(E/E_P)$ consistent with earlier literature. Among different choices in 
\cite{l2,l011} the choice $g_1=1$ and $g_2=1+\lambda E$ was considered where $\frac{1}{\lambda}\approx E_P$ but this choice leads to energy dependent theory of light. In \cite{l11,s} $g_1=g_2=\frac{1}{1+\lambda E}$ was considered which do not give a theory with varying $c$ and capable of solving the horizon problem. Here we will focus on the choice of \cite{l2} which considered $g_1 = \frac{1}{(1+\lambda_1 E)(1-\lambda E)}$ and $g_2=e^{\frac{E}{E_P}}$
with $\lambda_1 >> E_P^{-1}$. This choice among many others is argued to give a theory where relativity of inertial frames is preserved, all observers agree with the invariant energy scale $\lambda^{-1}=E_P$, the UHECR threshold can be increased and there will be a maximum momentum for the granularity of space. So with the choice $g_1=\frac{1}{(1+\lambda_1 E)(1- \lambda E)}$ and $g_2=e^{\frac{E}{E_P}}$ eqn. (\ref{wp1}) can be evaluated as
\begin{equation}
\Psi_{wp} = \frac{C_3}{2^{\nu +1}}~a^{q+1} ~\frac{e^{-\frac{a^{3(1-\alpha)}}{4B}}}{B^{\nu +1}}
\end{equation}
where $B=\xi +i \frac{3}{32}(1-\alpha)^2 T$ and $\xi = \frac{3(q+1)(1-\alpha)^2}{64E_P}$. Here we made the approximation $J_{\nu}\Big\{r e^{-\frac{9(1-\alpha)^3}{64 E_P}r^2} a^{\frac{3(1-\alpha)}{2}}\Big\} \approx J_{\nu}\Big\{r a^{\frac{3(1-\alpha)}{2}}\Big\}$ for the range of the integral. This approximation can be justified by the fact that $\lim_{E\rightarrow 0}g_2=1$. Using the many-worlds interpretation of quantum mechanics \cite{b28} we can calculate the expectation value of the scale factor as
\begin{equation}
\langle a \rangle (T) = \frac{\Gamma \big(\frac{5-3\alpha +2q}{3-3\alpha} \big)}{\Gamma \big(\frac{4-3\alpha +2q}{3-3\alpha} \big)} \left(\frac{\xi}{2}\right)^{\frac{1}{3(\alpha-1)}}\left[\xi^2 + \frac{9}{1024}(1-\alpha)^4~ T^2\right]^{\frac{1}{3(1-\alpha)}} ~~.
\end{equation}
So our model represents a bouncing universe with no singularity at $T=0$ and in the asymptotic limit $T\rightarrow \infty$ the model reduces to flat FRW universe.\par
This result is not at all new as we have just reproduced the result of \cite{iso}. But in \cite{iso} one has to consider an extra Gaussian superposition factor with strong intention for a finite norm wave packet. But here we see that a suitable choice of the Rainbow function $g_2$ allowed us to build a bouncing singularity free cosmological model which asymptotically behave like the flat FRW universe. Our result regarding the singularity is in good agreement with \cite{l} where the Rainbow modified FRW equations were treated as effective equations and compared the results with those of loop quantum cosmology. It is important to note that the expectation value of the scale factor depends on the operator ordering parameter of the Schr$\ddot{o}$dinger-Wheeler-deWitt equation $q$ which effects the bouncing nature of the model. As $\xi = \frac{3(q+1)(1-\alpha)^2}{64E_P}$ so we must avoid $q=-1$ to get a singularity free model. It is well known that the anisotropic models are plagued by the problem of nonunitarity \cite{aniso,nb}. So it may be interesting to study if some choice of the Rainbow functions can cure the nonunitarity in anisotropic models. So finally we would like to emphasize that it is possible to construct singularity free cosmological models in the formalism of Rainbow gravity with specific choices of the Rainbow functions $g_1$ and $g_2$. But it is very important to consider the approach with fundamental quantum fields which play essential role in the very early universe. 



\end{document}